\documentstyle[11pt]{article}
\def\a{\alpha}
\def\b{\beta}
\def\c{\gamma}\def\C{\Gamma}

\def\h{\eta}
\def\k{\kappa}

\def\m{\mu}

\def\s{\sigma}

\def\th{\theta}\def\Th{\Theta}

\def\beq{\begin{equation}}\def\eeq{\end{equation}}
\def\beqa{\begin{eqnarray}}\def\eeqa{\end{eqnarray}}
\def\barr{\begin{array}}\def\earr{\end{array}}

\def\del{\partial}
\def\ua{\underline{\alpha}}
\def\ub{\underline{\phantom{\alpha}}\!\!\!\beta}

\def\una{\underline a}\def\unA{\underline A}

\def\unc{\underline c}

\def\unM{\underline M}

\def\xz{\times}

\let\la=\label

\let\bm=\bibitem

\def\bd{\begin{document}}
\def\ed{\end{document}}
\def\ba{\begin{array}}
\def\ea{\end{array}}
\def\bea{\begin{eqnarray}}
\def\eea{\end{eqnarray}}
\def\ft#1#2{{\textstyle{{\scriptstyle #1}\over {\scriptstyle #2}}}}
\def\fft#1#2{{#1 \over #2}}
\newcommand{\be}{\begin{equation}}
\newcommand{\ee}{\end{equation}}

\begin{document}

\begin{titlepage}
\begin{flushright}
CERN-TH/96-200\\
King's College/kcl-th-96-??\\
Texas A \& M/CTP TAMU-28/96\\
ICTP/IC/96/126\\
hepth/9607227
\end{flushright}
\vskip 2cm
\begin{center}
{\Large{\bf Superbranes}}
\end{center}
\vskip 1.5cm
\centerline{\bf P.S. Howe \footnote[1]{Permanent address: Dept. of Mathematics,
King's College,  London}}
\vskip5mm
\centerline{CERN}
\centerline{Geneva, Switzerland}
\vskip 5mm
\centerline{and}
\vskip 5mm
\centerline{\bf E. Sezgin \footnote[2]{Permanent address: Center for
Theoretical Physics,
Texas A \& M University, College Station, Texas
77843}$^,$\footnote[3]{Research supported in
part by NSF Grant  PHY-9411543}}
\vskip 5mm
\centerline{International Center for Theoretical Physics, Trieste, Italy}

\vskip 1.5cm

\begin{abstract}

\noindent
We consider the doubly supersymmetric formulation of various $p$-branes, that
is,
we replace the worldsurface of a super $p$-brane by a super worldsurface
and  consider the
embedding of the latter into a target superspace. The number of worldsurface
fermionic
coordinates is taken to
be half as many as those of the target superspace. We show that a simple
geometrical constraint
and its integrability condition lead to manifestly worldsurface
supersymmetric field equations
for a large class of super $p$-branes. We illustrate this procedure in some
detail in the case
of  the $D=11$ superfivebrane. We also describe a
class of super
$p$-branes in which a worldsurface linear supermultiplet arises. In some
cases we show that an
additional constraint involving the curvature of an appropriate
worldsurface antisymmetric tensor
potential is needed to put the theory on-shell.

\end{abstract}

\vskip 2cm
\begin{flushleft}
CERN-TH/96-159\\
\today
\end{flushleft}
\end{titlepage}

\section{Introduction}

With the emergence of super $p$-branes as important ingredients in
non-perturbative string
physics, it has become more urgent to develop a better understanding of
their properties. For
example, still not much is known about the manifestly supersymmetric
actions that govern the
dynamics of these objects, with the exception of  Type I branes, by which
we mean those for
which the worldsurface degrees of freedom are described by a scalar
supermultiplet
\cite{hlp,bst,at}. It is now known that there are a number of other super
$p$-branes, called
Type II branes \cite{duff1}, in which the worldsurface degrees of freedom
form either Maxwell
multiplets (the D-branes in $D=10$ \cite{polch1}), or the tensor multiplet
(the $11D$
superfivebrane \cite{guven,gt}).

In all these cases it would be very useful to have manifestly supersymmetric
actions or, at least, equations of motion. Among other things, these would
help us  to gain a
better understanding of the duality symmetries of M-theory. Moreover, they
might help in
uncovering  various  relationships among the diversity of branes that exist
through the
examination of the properties of the relevant  brane actions or equations
of motion in
appropriate limits.

In the case of a superparticle or a superstring, it is well known that one
can either formulate
the action with manifest  worldline/worldsheet supersymmetry (NSR
formalism), or with manifest
target space supersymmetry (GS formalism). In some cases, there is also a
formulation in which
both the worldline/worldsheet and target space supersymmetry is manifest.
Variants of such
formulations are known as twistor, twistor-like or doubly supersymmetric
formulations. See, for
example, \cite{b1} for an extensive list of references. In fact, the
doubly-supersymmetric
approach is the main theme of this paper. However, before we present our
results, it is useful
to recall a few more facts about super $p$-brane actions.

To begin with, we recall that an NSR type formulation of $p$-branes beyond
$p=1$ does not seem to be possible, while a GS type formulation does exist,
at least for Type I
branes. The GS type formulation has a fermionic worldsurface symmetry, known as
$\k$-symmetry, which plays a crucial r\^{o}le in obtaining worldsurface
supersymmetry after gauge
fixing. It should be noted, however, that proving  worldsurface
supersymmetry is a very
nontrivial task which has been carried out fully only for superstrings
(see, for example,
\cite{gsw}) and the supermembrane in $D=4$ \cite{pkt2}.

As for D-branes, it is now well-known that they can be  described as
surfaces on which an open
superstring ends \cite{polch1}. This description makes it possible to
understand various
properties of their actions, in particular, the necessity of replacing the
usual Nambu-Goto action by the Dirac-Born-Infeld
generalization due to the presence of vector
fields.  However, it
should be noted that, so far, a $\k$-symmetric D-brane action is known only
for the Dirichlet
2-brane, and it is obtained from the $\k$-symmetric $D=11$ supermembrane
action by dimensional
reduction to $D=10$, followed by the dualization of the 11th scalar to a
world volume vector
\cite{pkt1}.

In the case of the eleven dimensional superfivebrane, since an open string
description is not
available, even the bosonic action is not known, although some partial
results have been
obtained \cite{pkt1,ah,eric1}. This is one of the most interesting cases,
and the case
which we will study in most detail in this paper.

The fact that Type II branes involve worldsurface supermultiplets other
than scalar multiplets
makes it difficult to apply the usual GS formalism. The requirement  of
$\k$-symmetry is very restrictive, and we recall that  worldsurface
supersymmetry is to be
expected only after gauge fixing. This state of affairs strongly suggests
that we should make
much greater use of worldsurface supersymmetry from the very beginning
while maintaining
manifest spacetime supersymmetry.

This brings us to the  main point of this paper, namely the description of
the dynamics of all
super $p$-branes, both Type I and Type II, by simple geometrical
considerations involving the
embedding of the world supersurface into the target superspace. The basic
idea is that once we
elevate the target superspace coordinates into worldsurface superfields, we
can see immediately
that there is room for the worldsurface supermultiplets in their
$\th$-expansions. In this
approach, $\k$-symmetry is traded for a more encompassing {\it local}
worldsurface
supersymmetry. The key issue is how to impose the right constraints on
these superfields. In
this paper we will show that, if we do not insist on an action to begin
with, but demand only
the right covariant equations of motion, the answer to this question is
very simple and is given
by the imposition of very natural embedding conditions that have clear and
simple geometrical
meanings. Once these constraints are well understood, one can of course
look for an action
principle from which they can be derived.

As mentioned earlier, the doubly supersymmetric aproach to super
$p$-branes, with emphasis on $p=0,1,2$, has been investigated previously in
many papers. The
basic idea was introduced in the context of  superparticles in $D=3,4$
\cite{stv,stvz}, and
extended to other superparticles in \cite{ht,ds,ghs,gs}, heterotic superstrings
in \cite{dghs,dis}, supermembranes in \cite{pt}, and higher super
$p$-branes in \cite{bs}. The
equations of motion of the supermembrane in $D=11$ were studied from the
point of view of
superspace embeddings in \cite{b1} and other Type I super $p$-branes were
studied in
\cite{b2} where a generalised action principle was used as a means of
deriving the corresponding
GS equations, although this is not an action principle in the usual sense
of the term. Moreover,
for $p\geq 2$, the worldsurface supermultiplets involved have not been
discussed in detail. One
of the main  results of this paper is that we clarify the precise
worldsurface multiplet
structures that arise due to the geometrical constraints in all possible
cases. The basic
embedding equation (1) (or its linearised version (\ref{mc})) in fact
determines various types of
worldsurface supermultiplet as a result of which we are able to say whether
the type of Lagrange
multiplier action first proposed in \cite{gs} is likely to succeed for the
brane in question or
not.

We shall begin by specifying this basic embedding condition and discussing
its geometrical
meaning.  We shall than fix a physical gauge to obtain a ``master
constraint equation'' that
encodes the equation of motion for the super $p$-brane in some cases.  We
will illustrate the
power of the master constraint by considering the interesting case of the
superfivebrane in
$D=11$, and show explicitly, at the linearised level, how the equations of
motion for the
worldsurface $d=6,N=(2,0)$ tensor multiplet emerge. We also mention the
worldsurface three-form
field strength and its relation to the target space four-form. This is the
superspace version of
the phenomenon observed in \cite{pkt1}.

We then go on to discuss the master equation for Type I branes in arbitrary
dimensions, D-branes
in $D=10$ and  what we call L-branes, which involve  worldsurface linear
multiplets. The L-branes
are new, but easy to understand: they correspond to dimensional reductions
of Type I branes by
one dimension, followed by dualization of the scalar corresponding to that
dimension to a
suitable worldsurface $p$-form potential. These are  analogous to the
Dirichlet twobrane
mentioned earlier, but they do not seem to have been considered before.

We shall also discuss the possibility of a superfivebrane in
$D=7$ which involves a worldsurface  $(1,0)$ tensor multiplet and comment
on a possible
ninebrane in $D=11$. In the conclusions we will comment on the nonlinear
aspects of the master
equation as well as a number of other aspects of our results. Throughout
the paper we restrict
our attention to target spaces with $N=1$ supersymmetry, except in the case
of $D=10$ where we
will consider both types of $N=2$ as well as $N=1$.

\section{The Embedding Equations}

The basic equation describing all branes as embedded subsupermanifolds is
simple to state and
has a simple geometrical meaning. Let $M$ denote the world-surface, $T$ its
tangent bundle,
$T_0$ and $T_1$ its even and odd tangent bundles, let $T^*$ etc denote the
corresponding dual
bundles, and use the same notation for the target superspace, but with
underlining. Let
$M\stackrel{f}\rightarrow\underline{M}$ denote the embedding of the
worldsurface in the target
superspace. The basic condition to be imposed on all embeddings we shall
consider is
\beq
T_1=T\cap \underline{T_1}\ , \la{t1}
\eeq
where on the RHS $\underline{T_1}$ is restricted to $M$. Dually one has
\beq
T_0^*=T^*\cap \underline{T_0}^* \ . \la{t2}
\eeq
The geometrical meaning of these statements is clear: the odd tangent space
of the worldsurface
at any point is a subspace of the odd tangent space of the target space at
the same point, and
the even cotangent space of the worldsurface at any point is a subspace of
the even cotangent
space of the target space at the same point. As we have mentioned, these
equations imply the
structure of the worldsurface multiplet as we shall show below. Several
cases can arise: the
world surface multiplet can be either on- or off-shell and one can have
either Type I or Type
II branes. If the representation is off-shell one can presumably construct
actions of the type
studied in \cite{gs,dghs}, whereas if it is on-shell the condition (\ref{t1})
above
determines the
worldsurface theory completely at the level of the equations of motion. In
the former case the
equations of motion are equaivalent to some additional geometrical
conditions which can be
stated using various $p$-forms on the target space and on the world-surface.

To justify these statements we introduce local bases $E_A=(E_a,E_\a)$ for
$T$ and $E^A=(E^a,E^{\a})$ for $T^*$ where we use the notation $A=(a,\a)$
for frame indices,
$M=(m,\m)$ for coordinate indices and where indices for $M$ and
$\underline{M}$ are distinguished by underlining the latter. We use latin
and greek indices to
refer to even and odd components respectively. We shall also use primed
indices to denote
quantities normal to $M$. We can set
\beq
E_A=E_A{}^{\unA} E_{\unA}\ , \la{ea}
\eeq
and
\beq
E^{\unA}=E^A E_A{}^{\unA}\ , \la{eua}
\eeq
where, in the latter equation, $E^{\unA}$ is pulled back onto $M$. In terms of
local coordinates $z^M$ and $\underline{z}^{\unM}$ we have
\be
E_A{}^{\unA}=E_A{}^M\del_{M}\underline{z}^{\unM}E_{\unM}{}^{\unA}
\ee
where $E_A{}^M$ is the inverse supervielbein on the worldsurface and
$E_{\unM}{}^{\unA}$ is the supervielbein on the target space.
The
basic equation
(\ref{t1}) translates to
\beq
E_{\a}{}^{\una}=0 \ . \la{me1}
\eeq
Dually, we have
\beq
E^{\una}=E^a E_a{}^{\una} \ . \la{de}
\eeq
The basic tensor on a supermanifold is the Frobenius tensor associated with
the odd tangent
bundle. Its components $T_{\a\b}{}^c$ with respect to a given basis are
defined by
\beq
\langle [E_\a,E_\b],E^c\rangle =-T_{\a\b}{}^c\ , \la{ft1}
\eeq
where $\langle,\rangle$ denotes the standard pairing between vectors and
forms. From
the above
definitions it is easy to deduce that
\beq
E_\a{}^{\ua}E_\b{}^{\ub}T_{\ua\ub}{}^{\unc}=T_{\a\b}{}^cE_c{}^{\unc}\ .
\la{me2}
\eeq
This equation is similar in some respects to the equation defining the
induced metric in
(pseudo)-Riemannian geometry. In (almost) all cases we shall consider one can
always
find a basis in
which the components of the  target space Frobenius tensor take the form
\beq
T_{\ua\ub}{}^{\unc}=-i(\C^{\unc})_{\ua\ub}\ . \la{ft2}
\eeq
The exceptional cases only differ from this form by the inclusion of invariant
tensors associated with internal symmetry groups.

To study the implications of these equations we consider two
simplifications. First, we shall
suppose that the target space is flat, and second, we shall take the
embedding to be
infinitesimal. If we denote the coordinates of $M$ by $(x^a,\th^{\a})$ and
those of $\unM$ by
$(X^{\una},\Th^{\ua})$, an infinitesimal embedding, in a standard gauge (
often referred to as
the physical gauge), is given by
\beq
\barr{lcllcl}
X^a &=&  x^a\ , \hspace{2cm} & X^{a'}&=&X^{a'}(x,\th)\ , \\
\Th^\a&=&\th^\a\ ,  & \Th^{\a'} &=& \Th^{\a'}(x,\th)\ .  \la{pg}
\earr
\eeq
(Recall that the primed indices refer to quantities normal to the world
supersurface.)
Substituting the above gauge choice into (\ref{me2}), and performing the shift
\beq
\tilde X^{a'}=X^{a'}-{i\over2}\th^{\a}(\C^{a'})_{\a\b'}\Th^{\b'}\ , \la{tx}
\eeq
we find that the linearised version of (\ref{me2}) is simply
\beq
D_\a \tilde X^{a'}=i(\C^{a'})_{\a\b'}\Th^{\b'}\ , \la{mc}
\eeq
where $D_\a$ is the flat superspace covariant derivative on the
worldsurface. It  satisfies
\beq
[D_\a,D_\b]=i(\C^a)_{\a\b}\ , \la{da}
\eeq
and the $\C$-matrices are those of the target space decomposed according to
the embedding. The leading components in the $\th$-expansion of
$\tilde{X}^{a'}$
and $\Th^{\a'}$ can be thought of as the Goldstone fields associated with the
partial breaking of translational symmetry and supersymmetry repectively.
Indeed
one could apply non-linear realisation theory using these fields, as has been
done for some cases previously \cite{hp,pkt3,ivanov}, but this is technically
complicated for all but the simplest cases whereas the current approach builds
in full covariance from the beginning.

Eq. (\ref{mc}) plays a central r\^{o}le in this paper and we shall refer to
it as
the ``master equation'' or ``master constraint''. One can now analyse the
various
possibilities that can arise according to the number of worldsurface
scalars and fermions that
are involved. We will enumerate these possibilities later. Examining these
possibilities, we
will learn that the master constraint (\ref{mc}) is sufficient to put the
theory on-shell in a
number of interesting cases, including the $D=11$ superfivebrane, in that it
describes correctly
the equations of motion for the relevant worldsurface supermultiplet. In
some other cases,
however, we will find that the master constraint (\ref{mc}) is {\it not}
sufficient to put the
theory on-shell. In such cases, we will propose a further constraint
involving appropriate
super $p$-form potentials on the world supersurface, whose r\^{o}le will be to
put the theory
on-shell.

The geometrical interpretation of these results, for the on-shell case, is
that, for Type I
branes the basic constraint corresponds to an adapted frame, that is, one
can consider
$E_\a{}^{\ua}$ to be part of a $Spin(1,D-1)$ matrix (up to a conformal
factor). In the Type II
case, the frame can no longer be considered as adapted because of the
appearance of extra
degrees of freedom associated with the worldsurface vector or tensor
multiplet. The appearance
of these extra degrees of freedom is related to actions of the
Dirac-Born-Infeld type which
occur in Type II. In the off-shell cases one also does not have the adapted
frame
interpretation although this is recovered on-shell in Type I.

We shall come back later to the points made briefly above. We now turn to
the analysis of the
master equation (\ref{mc}) in detail for the interesting case of the $D=11$
superfivebrane.

\section{The Eleven Dimensional Superfivebrane}

Let us consider equation (\ref{mc}) for the case of the superfivebrane in
$D=11$. It is convenient to introduce some notation which reflects the
$Spin(1,5)\xz USp(4)$ symmetry of six-dimensional superspace. A $D=11$ Majorana
spinor $\psi_{\ua}$ decomposes as
\beq
\psi_{\ua}=\big(\psi_{\a i},\psi^{\a}_i\big)\ , \la{gm}
\eeq
where $i=1,\ldots ,4$ is a $USp(4)$ index and $\a=1, \ldots ,4$ is a
six-dimensional Weyl spinor index with upper (lower) indices corresponding to
anti-chiral (chiral) spinors respectively. The six-dimensional spinors satisfy
a
$USp(4)$ symplectic Majorana-Weyl reality condition. One can choose a
representation for
the $D=11$ $\C$-matrices in which, for example,
\beq
(\C^a)_{\a i,\b j}=\h_{ij}(\s^a)_{\a\b}
\eeq
where $\h_{ij}$ is the $USp(4)$ antisymmetric invariant tensor and the $\s^a$
the
the six-dimensional chirally-projected gamma-matrices. Similarly the remaining
components are given
by products of six-dimensional and five-dimensional invariants.

The embedding equation (\ref{mc}) becomes
\beq
D_{\a i} \tilde X^{a'}=i(\c^{a'})_{ij} \Th_\a^j\ , \la{mca}
\eeq
where the $\c^{a'}$ are the five-dimensional (Euclidean) gamma-matrices
which are
antisymmetric. This equation defines the $d=6,\ N=(2,0)$ tensor multiplet
introduced in \cite{hst}. Its components consist of a self-dual third-rank
antisymmetric tensor, a set of chiral fermions transforming under the
four-dimensional representation of $USp(4)$ and five scalars. The scalars and
spinors are the leading components of $\tilde X^{a'}$ and $\Th_\a^i$
respectively, so it remains to locate the antisymmetric tensor and to show
that there are no further components. To do this one applies a spinorial
covariant derivative to (\ref{mca}) and uses the six-dimensional supersymmetry
algebra. One finds
\beq
D_{\a i} \Th_{\b j}=\h_{ij}L_{\a\b}-{1\over2}(\c^{a'})_{ij}(\s^a)_{\a\b}\del_a
\tilde X_{a'}\ . \la{dt}
\eeq
The symmetric bispinor $L_{\a\b}$ defines a self-dual third-rank antisymmetric
tensor as required. Continuing in this manner one finds by applying further
spinorial covariant derivatives that the fermion field satisfies the Dirac
equation, the scalar fields satisfy the Klein-Gordon equation and the tensor
field satisfies the Bianchi identity and field equation for a third-rank
antisymmetric field strength tensor. Furthermore, there are no other spacetime
components, so that equation (\ref{mc}) defines an on-shell tensor multiplet as
claimed.

For this case we have also verified that equation (\ref{mc}) is consistent
up to quadratic
order, and that it also implies that
\beq
T_{\a i,\b j}{}^c=-i\h_{ij}(\s^c)_{\a\b}\ , \la{torsion}
\eeq
up to a choice of basis on the world-surface. Furthermore, one can verify
that  the geometry
outlined above is consistent with the existence of a 3-form $H_3$ on $M$
such that
\beq
d H_3=\underline{H}_4\ , \la{hc}
\eeq
where $\underline{H}_4$ is the usual 4-form on $D=11$ superspace pulled back on
to the world supersurface. Note that it
is not necessary to impose this equation in order to obtain the required
worldsurface multiplet, as the embedding equation automatically ensures this.
However, equation (\ref{hc}) is very useful in the non-linear case as it
establishes
the existence of a potential for the antisymmetric tensor and shows how the
latter is related to the embedding tensor $E_\a{}^{\ua}$. A more detailed
exposition of the fivebrane in eleven dimensions will be given in a
forthcoming article.

\section{ The Superbrane Scan}

Given an embedded $p$-brane  world supersurface, the target space Lorentz
group breaks down to
$SO(1,p)\times SO(D-p-1)$. The transverse symmetry group $SO(D-p-1)\equiv
G^t$ will be identified
with the automorphism group $G^a$ of the worldsurface Popincar\'e
superalgebra, in all cases
except the Type I, codimension 4 embeddings (see later), where
the spin group corresponding to $G^t=Sp(1)\times Sp(1)'$, but
$G^a=Sp(1)$. The Goldstone superfield $X^{a'}$ always transforms under
the vector
representation of
$G^t$, while the Goldstino superfield $\Th^{\a'}$ and the worldsurface
fermionic coordinates
transform under equivalent or inequivalent fundamental spinor representations
of $SO(p,1)\times
G^a$, depending on the type of the embedding.

In determining the embedded world supersurface, we shall always maintain
the symmetries mentioned
above. This means that the Dirac matrices  $(\C^{a'})_{\a\b'}$ should exist
as invariant tensors
of these symmetry groups. While this requirement may sound rather innocent,
it restricts the
possible embeddings considerably, as we shall see below.

In our search for possible embeddings, we will restrict our attention to
minimal target
superspaces, with the exception of $D=10$, by which we mean $N=1$
superspaces in which the
number of fermionic coordinates is minimal. In $D=10$, we will consider the
$N=2$ cases as
well. To specify the relevant superspace more precisely, let us use the
notation
\be
           (D|D';\ SO(1,D-1)\ {\rm spinor\ type}\ )\ , \la{ts1}
\ee
where $D'$ is the real dimension of target space fermionic coordinates.
Thus, the minimal target superspaces we will consider are:
\be
\barr{lllll}
 (11|32;M)  & (10|16;MW) &(9|16;PM) &(8|16;PM) &(7|16;D) \\
 (6|8;W) & (5|8;D)   & (4|4;M)  & (3|2;M)  &
\earr
\la{ts2}
\ee
where $M$ stands for Majorana, $W$ for Weyl, $D$ for Dirac, $S$ for
symplectic and $P$ for
pesudo. For example, $SMW$ stands for symplectic-Majorana-Weyl, etc. In
$D=5,6,7$, we can equally well work with $PSM$, $SMW$ and $SM$ spinors,
respectively, in which
case the target Poincar\'e superalgebra acquires an $Sp(1)$ automorphism
group. The precise
definitions of these reality conditions can be found in \cite{kt}. See also
\cite{ss} for a
summary, in which the mostly positive spacetime signature $(-,+,+,...,+)$
adopted in this paper
is used. The analysis of the supermultiplets which  arise is done in each
case using the
method discussed in the previous section in the context of the $D=11$
fivebrane. We shall not
give any further details here; we shall simply state the results.

Let us now consider the embedding of Type I branes. In this class, both the
number of bosonic as
well as fermionic degrees of freedom are given by the codimension of the
embedding, which
for a $p$ brane in $D$ dimsnions is defined by $D-d=D-p-1$. The master
constraint for
codimension 8,4,1 embeddings takes the form
\be
D_{\a i}\,X^{a'} = i(\s^{a'})_{ij'}\,\Th_\a^{j'}\ ,
\quad\quad a'=1,...,D-p-1\ , \qquad \a=1,...,m\ , \qquad\ i,j'=1,...,n\ ,
\la{mc1}
\ee
where $(m,n)$ are the dimensions of the relevant $SO(1,p)\times G^a$
fundamental spinor, which
we will specify case by case below. The $\s$-matrices are the van der
Warden symbols, i.e.
chirally projected  $\c$-matrices, which obey the $SO(D-p-1)$ Clifford
algebra. The spinor
index $\a$ is also chirally projected when it labels a Weyl spinor. To list the
possibilities, let us use the following notation:
\be
(\ p,\,D,\, (m,n);\ SO(1,p)\times G^a\ {\rm spinor\ type}\ )\ .
\la{not}
\ee
The constraints (\ref{mc1}) are possible for the following (codimension
8,4,1) embeddings:
\be
 \barr{llll}
 {\rm codimension\ 8}: &(2,\,11,\,(2,8);\ M)  & (1,\,10,\, (1,8);\ MW) & \\
{\rm codimension\ 4}: &(5,\,10,\,(4,2);\ SMW) &(4,\,9,\,(4,2);\ PSM)  &
(3,\,8,\,(4,2);\ PSM)    \\
 & (2,\,7,\,(2,2);\ D) &(1,\,6,\,(1,2);\ W) &\\
 {\rm codimension\ 1}: & (2,\,4,\,(2,1);\ M)   & (1,\,3,\,(1,1);\ MW) &
 \earr
 \la{s1}
\ee

The codimension 8 cases are the supermembrane in $D=11$ and the heterotic
string in $D=1$;
the corresponding worldsurface supermultiplets are the on-shell $d=3,N=8$
scalar multiplet and
the off-shell $d=2,\ (8,0)$ scalar multiplet, respectively. The codimension
4 cases are the
fivebrane in $D=10, N=1$, the fourbrane in $D=9$, the threebrane in $D=8$,
the twobrane in
$D=7$ and the heterotic string in $D=6$. The corresponding worldsurface
multiplets are
hypermultiplets in dimensions $d=6,5,4,3$ and $2$ respectively, each such
multiplet having four
scalars. All of these multiplets are on-shell except for the string case. In
codimension 1, we have
the $D=4$ supermembrane
and the $D=3$ heterotic string. The corresponding worldsurface
supermultiplets are the
$d=3, N=1$ and $d=2,\ (1,0)$ scalar multiplets, respectively. Both of these
multiplets are
off-shell.

The remaining Type I branes have codimension 2, and the master constraint
for this case takes
the form
\be
D_\a X = \Th_\a~\la{mc2}\ ,
\ee
where we have defined the complex scalar $X = (X^{D-1}+i X^{D-2})/{\sqrt
2}$. This constraint is possible for the following (codimension 2) embeddings:
\be
 {\rm codimension\ 2}:\qquad (3,\,6,\,(2,1);\ W)\quad\quad  (2,\,5,\,(2,1);\ D)
\quad\quad
 (1,\,4,\,(1,1);\ W) \la{s2}
\ee

These branes are the threebrane in $D=6$, a twobrane in $D=5$ and the
heterotic string
in $D=4$, with corresponding worldsurface multiplets which are chiral
multiplets, with
$d=4,N=1$, $d=3, N=2$ and $d=2,\ (2,0)$ supersymmetry, respectively.

Next, we list the master constraints for $D$-branes.  As is well known,
Type IIA supergravity
has even branes with $p=2,4,6,8$, and Type IIB supergravity has  odd branes
with $p=3,5,7$. The
master constraints for the former case takes the form
\be
 D_{\a i}~X^{a'} = i(\c^{a'})_{ij}~\Th_\a^{j}\ ,
\quad\quad a'=1,...,D-p-1\ , \qquad \a=1,...,m\ , \qquad i,j=1,...,n\ ,
\la{mc3}
\ee
where the $\c$-matrices are $SO(D-p-1)$ matrices with $D-p-1=1,3,5,7$. More
precisely,
these constraints are possible for the following Type IIA branes:
\be
(8,\,10,\,(16,1);\ PM)\quad\quad (6,\,10,\,(8,2);\ SM)\quad\quad
(4,\,10,\,(4,4);\ PSM)\quad\quad
(2,\,10,\,(2,8);\ M) \ . \la{s3}
\ee
Note that the transverse symmetry group $G^t$ is identified with the
automorphism group
$G^a$, and that, unlike in the case of Type I branes, the world supersurface
fermionic
coordinates and the Goldstino fermions are in equivalent fundamental spinor
representations
of $G^a$.

For the twobrane, the fourbrane and the sixbrane the worldsurface multiplet
determined by the master equation is the maximally supersymmetric Maxwell
multiplet in $d=3,5$ and $7$ with $N=8,2$ and $1$
supersymmetry respectively. These multiplets are all on-shell. For the
eightbrane, which has codimension 1, the master equation gives an
unconstrained scalar superfield. An additional constraint of the type of
equation (\ref{hc}) will be necessary to put the theory on-shell in which
case the worldsurface mutliplet will be the $d=9$ Maxwell multiplet.

The master constraints for the Type IIB branes are one of the following
three types:

(i) the constraint (\ref{mc1})  for $(3,\,10,\,(2,4);\ W)$;

(ii) the constraint (\ref{mc1}) for $(5,\, 10,\,((4_+,2_+) +
(4_-,2_-));SMW)$ and

(iii) the constraint (\ref{mc2}) for $(7,\,10,\,(8,1);\ W)$\ .

In case (ii), $4_\pm$ refer to the left and right handed spinors of
$SO(5,1)$, and
$2_\pm$ refer to the left and right handed spinors of $G^a=SO(4)$. In this
case, the  constraint
(\ref{mc1}) is to be written for $(4_+,2_+)$ and $(4_-,2_-)$ separately.

For the Type IIB branes, the threebrane and the fivebrane again have
worldsurface multiplets
which are maximally supersymmetric Maxwell multiplets in $d=4$ and $d=6$,
that is, $N=4$ and
$N=(1,1)$, respectively. In the case of the  sevenbrane, however, the brane
has  codimension 2
and so the worldsurface supermultiplet is a chiral scalar superfield.
Moreover, unlike the case
of Type I codimension 2, this chiral multiplet cannot be used to write an
off-shell Lagrangian. A
further constraint, presumably of the type of equation (\ref{hc}), but with
a worldsurface
twoform, is necessary to put the theory on-shell. After this constraint has
been implemented, the
worldsurface multiplet will be the on-shell Maxwell multiplet in $d=8$.

We next examine the master constraints which correspond to a new class of
branes, which we
refer to as L-branes. The master constraint for these branes takes the form
given in
(\ref{mc3}), with the following  data:
\be
(5,\,9,\,(4,2);\ SMW) \quad\quad (4,\,8,\,(4,2);\ PSM) \quad\quad
(3,\,7,\,(4,2);\ M) \quad\quad
(3,\,5,\,(4,1);\ M) \ . \la{s4}
\ee
Note that the first three are codimension 3, and the last one is
codimension 1. Thus, the
transverse group is  $G^t=SO(3)$ for the first three cases, and the world
supersurface
fermions and the Goldstone spinor superfield carry the equivalent doublet
representation of
$SO(3)$. Note also that the L-branes,  for which $(p,D)$  is given by
$(3,5)$, $(3,7)$,
$(4,8)$ and $(5,9)$, correspond precisely to the double dimensional
reductions by one
dimension, followed by dualization of the scalar associated with this
dimension, of the
following  Type I branes: $(3,6)$ $(3,8)$, $(4,9)$ and $(5,10)$.

For the L-branes, the master equation determines the worldsurface mutiplet
to be the maximally
supersymmetric linear multiplet in all cases with codimension 3. For the
remaining case, which
has codimension 1, (i.e. the threebrane in $D=5$), the worldsurface
multiplet is an
unconstrained scalar multiplet. An additional constraint, presumably
involving antisymmetric
tensors, is necessary to obtain the $d=4, N=1$ linear multiplet. Note that
the linear
multiplets are always off-shell. Since the linear multiplets involve
antisymmetric tensor gauge
fields it seems probable that we will again have equations such as (\ref{hc}).

Finally, we point out two more possible master constraints. The first one
is a fivebrane in
$D=7$, for which the resulting world supersurface multiplet is a $d=6$
tensor multiplet. This
constraint is
\be
D_{\a i}~X = \Th_{\a i}\ ,\quad\quad  \a=1,...,4 \ , \qquad i=1,2\ , \la{l2}
\ee
where the $\th^{\a i}$ are $SMW$ spinors in six dimensions. We expect that
this constraint,
which by itself gives rise to a scalar superfield, will, when supplemented
by a suitable
$H$-constraint, describe the $(1,0)$ tensor multiplet in $d=6$. This latter
multiplet is
on-shell and involves a self-dual third-rank antisymmetric tensor.

The second possibility is a ninebrane in $D=11$. The master equation takes
the simple form
$D_\a X= \Th_\a$, where the spinor is 16 component Majorana-Weyl, thereby
leading to a real
scalar superfield $X$ in $d=10, N=1$. This is reducible, and could lead to
a spin 3/2 multiplet
with an additional constraint, or perhaps to a spin 2 multiplet. However,
it is not clear that
either of these multiplets would make sense in the interacting case and
neither is it clear
which additional constraints would need to be imposed.

\section{Concluding Remarks}

To summarise the preceeding sections, we have shown that the embedding
equation (\ref{me1}), or
more precisely its linearised form (\ref{mc}), determines various
worldsurface supermultiplets.
These supermultiplets can be on-shell, off-shell irreducible or off-shell
reducible. In the
first case we arrive directly at the equations of motion while in the other
two cases
additional constraints are needed to get irreducibility and/or the
equations of motion. We
expect that these additional constraints should involve various forms on
both the target space
and the worldsurface. For Type II branes and L-branes these equations
should be of the type
given in equation (\ref{hc}) while in the Type I case an appropriately
modified pull-back of
the target space $(p+1)$-form field strength should vanish, as discussed
for various cases in
\cite{dghs,b1,b2,bs}.

The main focus of attention in this paper has been the linearised brane
equation (\ref{mc}) and
the resulting supermultiplets. However, the full non-linear theory can be
studied using
equations (\ref{me1}) and (\ref{me2}) as a basis. Moreover, non-trivial
target spaces can be
incorporated straightforwardly. The method should therefore be powerful
enough to determine the
full non-linear structure of the eleven-dimensional fivebrane, at least at
the level of the
equations of motion. This will be the subject of a forthcoming article. We
note here that, in
the case of Type II branes, the departure of the embedding from involving a
straightforward
adapted frame is a reflection of the Dirac-Born-Infeld structure of the
bosonic action. In
addition, the induced geometry on the worldsurface will play a more
important r\^{o}le in the
non-linear theory. Our expectation is that in many cases this geometry will
correspond to
off-shell conformal supergravity. This is not unreasonable since the
induced supergravity
multiplet is composite, and hence off-shell, even when the equations of
motion of the brane are
satisfied.

We have commented earlier on the difficulties that have been encountered in
constructing doubly
supersymmetric actions for various branes. The analysis we have given shows
clearly that, for
those branes for which the master constraint leads to on-shell multiplets,
actions involving a
Lagrange multiplier to impose this constraint will not work; one will
inevitably encounter
additional unwanted propagating degrees of freedom. For those branes which
have irreducible
off-shell worldsurface multiplets, on the other hand, one might hope that
such actions might
work. We note that our list includes all the examples of this type which
have already been
found but that there are several new possibilities\footnote{The $D=3,N=2$
string considered in \cite{gs2} is a double dimensional reduction of the $D=4,
N=1$ supermembrane}. We further remark that
finding actions for
many of the on-shell branes looks to be a difficult task. For example, the
Type II branes in
$D=10$ have worldsurface multiplets which are maximally supersymmetric
Maxwell multiplets for
which no off-shell extensions (leading to Lagrangians) are known. In the
case of fivebranes
in $D=11$ and $D=7$ the world surface multiplets involve self-dual
antisymmetric tensors, and
it is well-known that there are severe problems in formulating Lorentz
covariant actions for
such objects even before one takes off-shell supersymmetry into account. On
the other hand, we
note that, leaving aside the $D=6$ heterotic string for which the
wordsurface supermultiplet is off-shell, the codimension 4, Type I branes
all involve hypermultiplets in
various guises. The
hypermultiplet is an on-shell multiplet but in this case it is known how to
go off-shell by
using harmonic superspace. We might therefore conjecture that actions for
these branes could be
found by enlarging the worldsurface to include the harmonic variables.

The branes we have considered here correspond in many cases to soliton
solutions of supergravity theories. It would therefore be interesting to see if
there are solitons related to the proposed L-branes and the $D=7$ fivebrane.
(In
the latter case, there is a candidate soliton \cite{dilaton}, which arises
in the gauged
$D=7, N=1$ supergravity with a topological mass term.) In particular, it
would be interesting
to see if one could approach this topic from a superspace perspective as it
might help one to
understand the close relationship between the target space and worldsurface
points of view. A
related topic is the question of branes with less than one-half
supersymmetry. In particular,
it would be interesting to see how the intersecting branes \cite{guvenv} found
in $D=11$
would fit into the
geometrical picture advocated here.

\vskip 2mm

E.S.~would like to thank the International Center for Theoretical Physics
in Trieste
for hospitality.

\pagebreak

\end{document}